\documentclass[sigconf]{acmart}

\AtBeginDocument{%
  }

\setcopyright{acmcopyright}
\copyrightyear{2023}
\acmYear{2023}

\acmPrice{15.00}
\acmISBN{978-1-4503-XXXX-X/18/06}

\usepackage{url}            
\usepackage{booktabs}       
\usepackage{amsmath}
\usepackage{amsfonts}       
\usepackage{nicefrac}       
\usepackage{microtype}      
\usepackage{xcolor}         

\usepackage{dblfloatfix}
\usepackage{graphicx}
\usepackage{float}
\usepackage{xspace}

\usepackage{amssymb}

\usepackage{colortbl}
\usepackage{multirow}
\usepackage{tabulary}
\usepackage{etoolbox}

\usepackage{algorithm}
\usepackage[noend]{algpseudocode} 
\usepackage{xspace}
\usepackage[breakable]{tcolorbox}
\usepackage{todonotes}
\usepackage{paralist}
\usepackage{tikz}
\usepackage{pgfpages}
\usepackage{pgfplots}
\usepackage{pgfplotstable}
\usepackage{subcaption}
\usepackage{tabularx}

\newcounter{myremark}

{\bfseries}{\rmfamily}

\newcommand{\tool}{\textsc{Mosaic}\xspace}

\newcommand{\myparagraph}[1]{\vspace{0.5em}\noindent{\bf #1.}}
\newtheorem{mydefinition}{Definition}{\bfseries}{\normalfont}

\newcommand{\matlab}{\texttt{MATLAB}\xspace}
\newcommand{\simulink}{\texttt{Simulink}\xspace}
\newcommand{\breach}{\textsc{Breach}\xspace}

\newcommand{\prism}{\textsc{Prism}\xspace}

\newcommand{\acc}{\texttt{ACC}\xspace}

\newcommand{\afc}{\texttt{AFC}\xspace}

\newcommand{\cstr}{\texttt{CSTR}\xspace}

\newcommand{\ddpg}{\texttt{DDPG}\xspace}
\newcommand{\tdthr}{\texttt{TD3}\xspace}
\newcommand{\sac}{\texttt{SAC}\xspace}

\newcommand{\NN}{\ensuremath{\mathsf{N}}\xspace}
\newcommand{\plant}{\ensuremath{\mathcal{R}}\xspace}

\newcommand{\af}{\ensuremath{\mathtt{AF}}\xspace}
\newcommand{\afref}{\ensuremath{\mathtt{AFref}}\xspace}

\newcommand{\drel}{\ensuremath{\mathtt{D_{rel}}}\xspace}
\newcommand{\vego}{\ensuremath{\mathtt{v_{ego}}}\xspace}
\newcommand{\dsafe}{\ensuremath{\mathtt{D_{safe}}}\xspace}
\newcommand{\vtarget}{\ensuremath{\mathtt{v_{target}}}\xspace}

\newcommand{\mdp}{\ensuremath{\mathcal{M}}\xspace}
\newcommand{\mdpstate}{\ensuremath{\mathsf{S}}\xspace}
\newcommand{\mdpinitalstate}{$s_0$\xspace}
\newcommand{\action}{\ensuremath{\mathsf{Act}}\xspace}
\newcommand{\prob}{\ensuremath{\delta\xspace}}
\newcommand{\ap}{\ensuremath{\mathsf{AP}}\xspace}
\newcommand{\mdplabel}{\ensuremath{\mathsf{L}}\xspace}

\newcommand{\mdptransition}{\ensuremath{\Theta\xspace}}

\newcommand{\pctlUntilOp}[1]{\mathbin{\mathcal{U}^{#1}}}
\newcommand{\NextOp}{\mathcal{X}\xspace}
\newcommand{\pctlProbOp}[1]{\mathbin{\mathcal{P}_{#1}}}

\newcommand{\moore}{\ensuremath{\mathcal{X}}\xspace}
\newcommand{\moorestate}{\ensuremath{\mathsf{Q}}\xspace}
\newcommand{\mooreinitialstate}{$q_0$\xspace}
\newcommand{\mooreinalphabet}{\ensuremath{\Sigma}\xspace}
\newcommand{\mooreoutalphabet}{\ensuremath{\mathsf{O}}\xspace}
\newcommand{\mooretransition}{\ensuremath{\Xi}\xspace}
\newcommand{\mooreoutput}{\ensuremath{\mathsf{G}}\xspace}

\newcommand{\actabst}{$\zeta_{\action}$\xspace}
\newcommand{\labelabst}{$\zeta_{\mdplabel}$\xspace}
\newcommand{\stateabst}{$\zeta_{\mdpstate}$\xspace}
\newcommand{\transabst}{$\zeta_{\mdptransition}$\xspace}

\newcommand{\Defeq}{:=}

\newcommand{\Robust}[2]{{ \llbracket #1, #2 \rrbracket}}
\newcommand{\UntilOp}[1]{\mathbin{\mathcal{U}_{#1}}}
\newcommand{\signal}{\ensuremath{r}\xspace}
\newcommand{\STLspec}[2]{\ensuremath{\varphi_{#1}^{#2}}\xspace}
\newcommand{\queue}{\ensuremath{\mathrm{Q}}\xspace}
\newcommand{\localOpt}{\ensuremath{\textsc{Opt}}\xspace}
\newcommand{\flag}[1]{\ensuremath{\mathtt{flag}_{#1}}\xspace}
\newcommand{\rb}[1]{{\mathtt{rb}_{#1}}}
\newcommand{\falRand}{\texttt{Random}\xspace}
\newcommand{\toolRand}{\tool_{rand}\xspace}

\begin{document}

\title{\tool: Model-based Safety Analysis Framework for\\ AI-enabled Cyber-Physical Systems}


\author{Xuan Xie$^1$, Jiayang Song$^1$, Zhehua Zhou$^1$, Fuyuan Zhang$^2$, Lei Ma$^{1,3}$}
\affiliation{%
  \institution{$^1${University of Alberta, Canada}\quad $^2${Kyushu University, Japan}  \quad$^3${The University of Tokyo, Japan}\\
  xxie9@ualberta.ca, jiayan13@ualberta.ca, zhehua1@ualberta.ca, fuyuanzhang@163.com, ma.lei@acm.org}
\country{}
 }








\renewcommand{\shortauthors}{X. Xie, J. Song, Z. Zhou, F. Zhang, and L. Ma}

\begin{abstract}
Cyber-physical systems (CPSs) are now widely deployed in many industrial domains, e.g., manufacturing systems and autonomous vehicles. 
To further enhance the capability and applicability of CPSs, there comes a recent trend from both academia and industry to utilize learning-based AI controllers for the system control process, resulting in an emerging class of AI-enabled cyber-physical systems (AI-CPSs). 
Although such AI-CPSs could achieve obvious performance enhancement from the lens of some key industrial requirement indicators, due to the random exploration nature and lack of systematic explanations for their behavior, such AI-based techniques also bring uncertainties and safety risks to the controlled system, posing an urgent need for effective safety analysis techniques for AI-CPSs. 
Hence in this work, we propose \tool, a model-based safety analysis framework for AI-CPSs. 
\tool first constructs a Markov decision process (MDP) model as an abstract model of the AI-CPS, which tries to characterize the behaviors of the original AI-CPS.
Then, based on the derived abstract model, safety analysis is designed in two aspects: online safety monitoring and offline model-guided falsification. 
The usefulness of \tool is evaluated on diverse and representative industry-level AI-CPSs, the results of which 
demonstrate that \tool is effective in providing safety monitoring to AI-CPSs and enables to outperform the state-of-the-art falsification techniques, providing the basis for advanced safety analysis of AI-CPSs.


\end{abstract}





\maketitle

\section{Introduction}

\begin{figure}[!t]
\centering
\includegraphics[width=0.75\columnwidth]{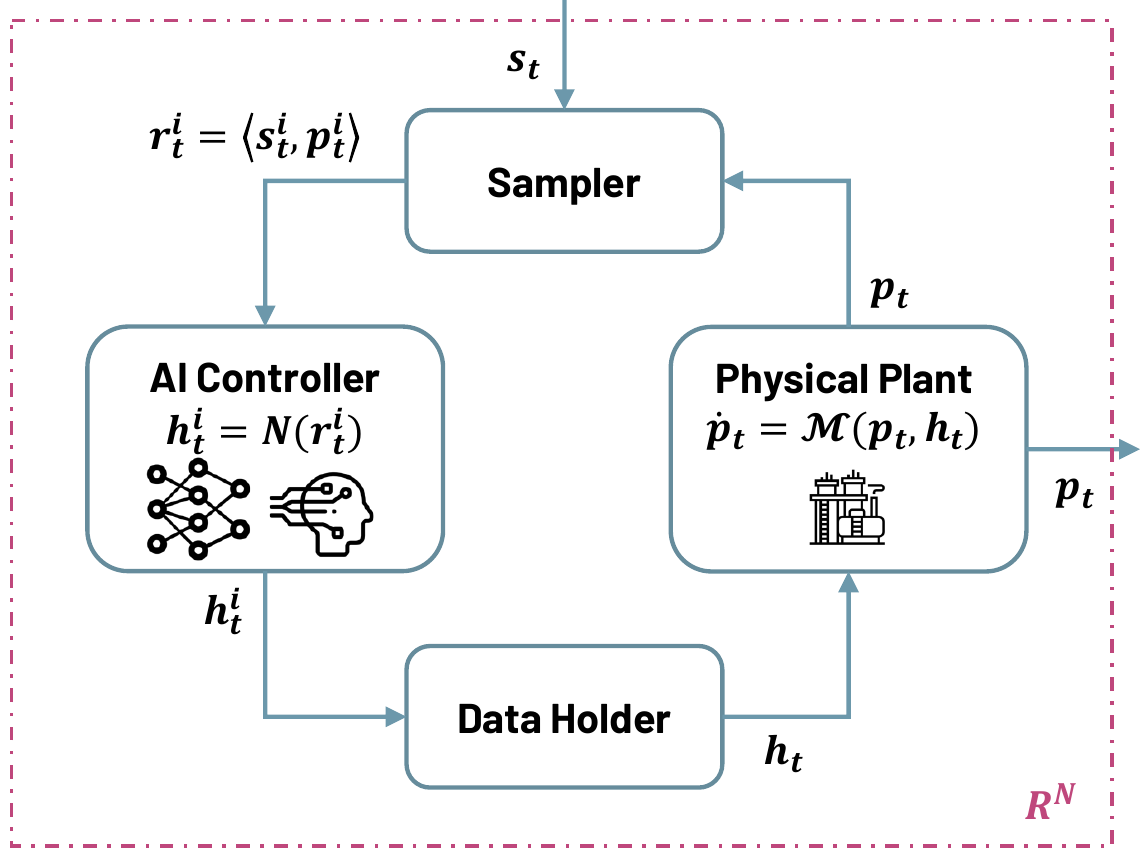}
\caption{The common workflow of AI-CPS }
\vspace{-10pt}
\label{fig:aicps}
\end{figure}

Cyber-physical systems (CPSs) are commonly and broadly defined as systems that integrate digital computational components and physical components (plants).
Benefiting from modern advances in digitization over the past decade, nowadays CPSs have been widely deployed and have become pillars of vast crucial industrial and social infrastructures across domains, e.g., industrial manufacturing systems~\cite{CSTRMathworks}, robotic systems~\cite{deshmukh2017testing}, computerized vehicle and aircraft controls~\cite{jin2014powertrain,ACCMathworks,heidlauf2018verification}, smart grids~\cite{stanovich2013development}, medical devices~\cite{lee2011challenges}, etc.
Unlike regular control systems or embedded systems, the ability to communicate between digital devices and physical processes enables CPSs to accomplish complicated tasks.
However, it is also common that uncertainties and difficulties in the controller design become the key challenges to ensuring the safety of CPSs~\cite{zhang2016understanding}.
Hence, how to realize an efficient and reliable control process of CPSs remains a research challenge.


Inspired by the impressive performance of using learning-based artificial intelligence (AI) techniques to solve intricate real-world problems, e.g., image recognition~\cite{alom2018history,iandola2016squeezenet} and decision-making~\cite{li2017deep,mnih2013playing}, there comes a recent trend of investigating the possibility of employing AI-based approaches to further enhance the control process of CPSs.
Such an AI-enabled CPS (AI-CPS) often demonstrates higher levels of performance and advantage in various aspects, such as efficiency, effectiveness, adaptability, and survivability, 
e.g., exhibiting better autonomy and intelligence.
Compared to traditional CPSs, AI-CPSs, in general, are able to work with complex system structures as well as complicated tasks and environments.
However, limited by the data-driven or random exploration nature of the learning-based AI techniques, an obvious potential drawback and risk of AI-CPSs is the lack of promising safety guarantees~\cite{Song2021WhenCS,zhang2022falsifai}, which poses concerns for wider adoptions, especially in safety-critical domains.  
To this end, a general technique and framework for safety analysis of AI-CPSs are highly desirable, which can be the foundation for building safe and trustworthy AI-CPSs.



The traditional and \emph{de facto} way of analyzing the safety of CPSs often relies on expert experience and a transparent understanding of the system behavior~\cite{lyu2019safety}. 
Many safety analysis methods have been developed to tackle the safety challenges of CPSs, such as fault tree analysis (FTA)~\cite{dafflon2021challenges}, failure modes and effects analysis (FEMA)~\cite{ebeling2019introduction}, model-based engineering (MBE)~\cite{banerjee2011ensuring}, etc. 
However, these techniques are in general not applicable to AI-CPSs due to reasons such as the low explainability of AI components, the limited testing samples, and the lack of a comprehensive model to describe system characteristics.
Moreover, 
existing falsification techniques designed for traditional CPSs are also still limited in falsifying and detecting safety issues of AI-CPSs~\cite{Song2021WhenCS}. 
Therefore, new safety analysis methods that are able to reveal the behavior of AI components are urgently needed to realize a safe and reliable deployment of AI-CPSs.



In this paper, we propose a model-based safety analysis framework, named \tool, for AI-CPS, which is mainly composed of three key parts: \emph{data collection}, \emph{model abstraction}, and \emph{safety analysis} (see Fig.~\ref{fig:overview}).
The central idea is that, through using simulation data that represents the safety properties of the AI-CPS under analysis, we first construct a Markov decision process (MDP)~ \cite{puterman1990markov} model as an abstract model of the system.
Such an abstract model possesses reduced state, input and output spaces and therefore makes an efficient safety analysis process possible. 
Then, based on the constructed abstract MDP model as the foundation, we propose safety analysis techniques for AI-CPSs from two directions: (1) online safety monitoring by utilizing probabilistic model checking (PMC)~\cite{kwiatkowska2011prism}; and (2) offline model-guided falsification.
A more detailed overview of \tool is presented in Section~\ref{subsec:overview}.
To demonstrate the usefulness of our proposed framework, we perform an in-depth evaluation on diverse and representative AI-CPSs across domains. 
The results demonstrate that \tool is effective and efficient in providing safety analysis to AI-CPSs, providing the basis for developing more advanced AI-CPSs.

In summary, the key contributions of this work are as follows:
\begin{compactitem}[$\bullet$]
    \item We propose a novel model-based safety analysis framework for AI-CPS, based on abstraction and refinement.
    The constructed abstract model is facilitated by probabilistic nature and safety awareness, which portrays the safety-related behavior of the system.

    \item Based on the abstract model, we propose two
    techniques for safety analysis of AI-CPS, i.e., (1) online safety monitoring and (2) offline abstract model-guided falsification.

    \item The online model-based safety monitoring technique enables to provide the safety advice online to avoid the potentially risky behaviors and hazards of AI-CPSs.
    
    \item Our offline model-guided falsification technique, which leverages the combined global model-guided exploration and local optimization-driven search-based failure input detection,
    performs offline safety analysis of an AI-CPS to detect the potential risky cases that could trigger the violation of safety conditions.

    \item The effectiveness and usefulness of \tool are demonstrated by our in-depth evaluation and analysis from three perspectives.
    (1) We show that our constructed abstract model is accurate in terms of state labeling, which is essential for further safety analysis. 
    (2) We also present that online safety monitoring can increase the safety of the system while maintaining its performance, which indicates its effectiveness. 
    (3) For falsification, we demonstrate that \tool outperforms three state-of-the-art falsification techniques on multiple diverse and representative AI-CPSs.
    The evaluation results are consistent with our expectations and confirm the usefulness of \tool as the basis for the safety analysis of AI-CPS. 
    
\end{compactitem}

To the best of our knowledge, this paper is a very early work that makes special focuses on the safety analysis of AI-CPSs. 
Witnessing the increasing trend of attempting to adopt AI to CPS, although AI could empower CPS to enhance the performance from various angles, on the other side of the double-edged sword, the potential risks and hazards (especially brought by issues such as uncertainty, behavior interpretability) post big concerns that potentially hinder the more widespread adoption of such AI-CPSs. 
At an early stage, we believe that building a safety analysis framework for AI-CPS could be a very important step, (1) not only in providing the technique and tool for safety issue analysis and detection, (2) but also facilitating further research along this important direction as the basis, towards designing more advanced safety and trustworthiness techniques to enable more widespread adoption of AI-CPSs.

To enable further research studies in this direction, we make all of our source code, benchmarks, and detailed evaluation results publicly available at {\url{https://sites.google.com/view/ai-cps-mosaic}.

\section{Overview}
\label{sec:overview}

\begin{figure*}[!t]
\centering
\includegraphics[width=\textwidth]{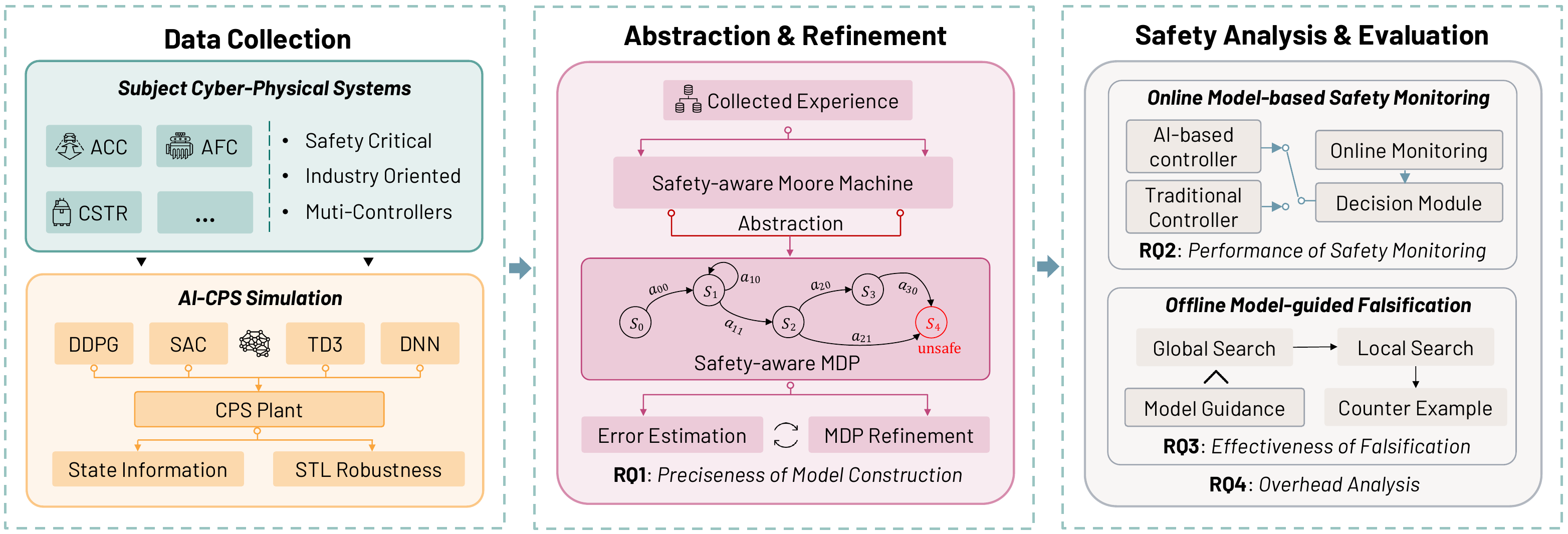}
\caption{Overview of \tool: a model-based safety analysis framework for AI-CPS}
\vspace{-10pt}
\label{fig:overview}
\end{figure*}

In this section, we first present a brief introduction to the structure of AI-CPS considered in this work. 
Then, an overview of the proposed safety analysis framework is given, together with the high-level research questions (RQs) that we would investigate.  

\subsection{AI-CPS}\label{subsec:aicps}
In general, an AI-CPS $\plant^\NN$ combines a physical plant with electrical devices, and through the communication between the powerful AI-based controller and the plant, which achieves highly competitive performance. 
As depicted in Fig.~\ref{fig:aicps}, it consists of four components: a sampler, a plant $\plant$, a data holder, and an AI controller $\NN$.
The sampler combines the external signal $s_t$ and the state of the plant $p_t$ as its output data $r^i_t = \langle s^i_t, p^i_t \rangle$, where $i$ indicates the time index.
Then, the AI controller takes the sample data $r^i_t$ as the input and determines an action $h^i_t$, which is sent to the data holder for producing a continuous signal $h_t$ to control the plant.
Based on the received control commands and signals, the plant makes the corresponding physical process (e.g., interactions with the environment) and evolves to new states to gradually complete the task. 
From the system perspective, the AI-CPS $\plant^\NN$ can be considered as a black box function that describes the physical process and
maps a system input $s$ to a system output $\plant^\NN(s)$.
A typical example of AI-CPS is the adaptive cruise control system~\cite{Song2021WhenCS,zhang2022falsifai,ACCMathworks}, which aims to control the ego car to keep a target velocity while avoiding collision with a lead car.
In this system, the sampler receives external signals, i.e., the velocity of the lead car and the distance between the two cars, as the input and sends it to the controller.
The AI controller decides the acceleration of the ego car, and passes the actions to the data holder.
The data holder composes the discrete actions as continuous signals and sends them to the plant for execution.

It is worth mentioning that, to achieve learning-based AI controllers, state-of-the-art techniques mostly fall into the following two categories: \emph{supervised learning} and \emph{deep reinforcement learning (DRL)}.
In supervised learning, training data is collected from traditional control-theoretical controllers, e.g., proportional-integral-derivative (PID) controllers and model predictive controllers (MPC), and used for learning the control policy.
In DRL, the AI controller is learned by directly interacting with the environment (by reinforcement learning) toward achieving optimal policy that maximizes a predefined reward function based on the observed system states.





\subsection{Overview of \tool and RQ Design}
\label{subsec:overview}

Figure~\ref{fig:overview} presents the workflow of \tool, which predominantly contains three key parts: \emph{data collection}, \emph{model abstraction}, and \emph{safety analysis}.

First, as the preparation step, we simulate the AI-CPS under analysis and collect relevant data that includes the states and traces of the system, as well as their safety properties.
Then, the collected data is used to build a Moore machine~\cite{kohavi2009switching} that provides a suitable representation of the behaviors of the AI-CPS for further safety analysis. 
In practice, the state, input, and output representation spaces of such a Moore machine are often high-dimensional and continuous, which poses computational challenges in performing the safety analysis.
Therefore, to address such a challenge, we propose constructing an abstract model from the Moore machine as an MDP by using a four-level abstraction in terms of state, transition, action and labeling.
While preserving the representativeness of critical safety properties, such an abstract MDP model enables an efficient analysis for AI-CPSs.
Since the preciseness of the constructed abstract model is imperative for performing further analysis,
therefore, the first research question that we would like to investigate is,
\textbf{RQ1:} \textit{How precise are the constructed abstract models?}

Based on the constructed abstract MDP model, we further propose safety analysis techniques from two directions: \emph{online safety monitoring} and \emph{offline falsification}.
As the first direction, we propose an online safety monitoring method that aims to increase the safety of the system while maintaining a similar performance of the original system.
In particular, the monitoring module intelligently computes online safety predictions by observing the system status and performing PMC on the abstract MDP model.
Then, according to the safety predictions, the actually applied controller is switched between the efficient AI-based controller and a predefined safety controller for keeping the AI-CPS safe.
For examining, whether online safety monitoring is able to result in a safety improvement of AI-CPS while keeping a similar performance compared to the original system, we would like to investigate,
\textbf{RQ2:} \textit{Can \tool provide effective safety monitoring?}

As the second direction, we further propose a novel offline model-guided falsification technique specially designed for AI-CPS.
Falsification is a well-established safety validation technique that explores the CPS system behavior space to search for a counterexample that violates the specification.
However, traditional falsification is ineffective in AI-CPS since it easily falls into the local optimum~\cite{Song2021WhenCS}.
To address this problem, we design and develop a novel falsification technique that combines global model-guided search and local optimization-based search to effectively detect counterexamples for AI-CPS.
To assess whether the proposed technique is useful and outperforms existing state-of-the-art falsification techniques for AI-CPSs, we perform a comparative study to demonstrate,
\textbf{RQ3:} \textit{Is \tool effective in guiding the falsification procedure?}

Furthermore, as an overall analysis, we would like to investigate how much overhead is introduced by safety monitoring.
Hence, we record the time cost by the query and analyze its impact on the online safety monitoring process to examine the ratio of the time spent on the monitoring components and the whole simulation.
This leads to another RQ that we would like to investigate,
\textbf{RQ4:} \textit{How much overhead is introduced by the safety query in the analysis?}

\section{Abstract Model Construction}
\label{sec:MDPcons&ref}



In this section, we discuss how to construct an MDP model as an abstract model of the AI-CPS to enable safety analysis.
We first briefly introduce how to use a Moore machine to capture the behavior of AI-CPS in Section~\ref{subsec:aicpsasmoore}.
Then, we propose a method to construct a representative abstract MDP model from collected data based on a four-level abstraction (Section~\ref{subsec:MDPcons}).
Finally, a model refinement technique is proposed to reduce the estimation error of the obtained abstract model (Section~\ref{subsec:MDPref}). 


\subsection{AI-CPS as Moore Machine}
\label{subsec:aicpsasmoore}

In general, the exact behavior of AI-CPS can be modeled as a \emph{Moore machine}~\cite{moore1956gedanken} that is defined as follows.

\begin{mydefinition}[Moore Machine]
A Moore machine \moore is a tuple (\moorestate, \mooreinitialstate, \mooreinalphabet, \mooreoutalphabet, \mooretransition, \mooreoutput), where \moorestate is a finite set of system states and \mooreinitialstate $\in \moorestate$ is the initial state. 
\mooreinalphabet and \mooreoutalphabet are finite sets and are referred to as input and output alphabets, respectively. 
\mooretransition $:\moorestate \times \mooreinalphabet \rightarrow \moorestate$ is the transition function that maps a state and the input alphabet to the next state.
\mooreoutput $:\moorestate \rightarrow \mooreoutalphabet$ is the output function mapping a state to the output alphabet.
\end{mydefinition}

More concretely, the behavior of AI-CPS is mapped to the components of Moore machine as follows.
\moorestate represents the system state space, and \mooreinitialstate is the starting point.
\mooreinalphabet represents the output space of the controller, which decides how the system behaves in the environment.
For \mooreoutalphabet, we use the robust semantics of signal temporal logic (STL), which is explained in the next paragraph, as the output of the system state.
\mooretransition describes how the status of the system changes after each control action, and \mooreoutput maps the status of the system to the robust semantics.





\myparagraph{Signal Temporal Logic}
With the ability to describe safety-related temporal behaviors, STL is extensively employed as the specification language of CPSs.
STL utilizes quantitative \emph{robust semantics}, which represents the degree of satisfaction of a certain specification as a quantitative value. 
Given an STL specification $\varphi$ and a system output signal $\plant^\NN(\signal)$ that corresponds to an input signal \signal, the STL semantics is a function $\Robust{\plant(\signal)}{\varphi}$ that maps $\plant(\signal)$ and $\varphi$ to a real number, which is denoted as the \emph{robustness} of $\plant(\signal)$ w.r.t. $\varphi$. 
This value indicates how robust $\plant(\signal)$ satisfies $\varphi$: the larger the value, the stronger the robustness of $\plant(\signal)$ satisfying $\varphi$. 
And if the value is negative, then it means that $\plant(\signal)$ violates $\varphi$. 
An example of an STL safety specification for the adaptive cruise control system (recall Section~\ref{subsec:aicps}) is $\Box_{[0,30]}(\mathtt{speed} \le 60)$, which stands for the speed of the ego car should not exceed 60 km/h within 30 seconds.
If the robust semantics returns $-2$, it means that the system violates the given speed requirement. 

In this work, we focus on the safety analysis of AI-CPS, which can be well described by using STL robust semantics as the degree of satisfaction of the system w.r.t. given safety specifications.
Therefore, we leverage the quantitative robust semantics of the STL specification as the output \mooreoutput of the Moore machine.

\subsection{MDP Model Construction}\label{subsec:MDPcons}

In practice, it can usually be computationally expensive to perform safety analysis directly on the Moore machine model of AI-CPS, since, on the one hand, the exact knowledge about model parameters is unknown.
On the other hand, the state, input, and output spaces of the Moore machine are often high-dimensional and continuous, causing computational challenges.
Nevertheless, we are still able to simulate the AI-CPS for collecting data that represents the behavior of the Moore machine, which includes the states and traces of the system, the actions of the AI controller, as well as the degree of satisfaction of the system w.r.t. given specifications.

By using the collected simulation data, we propose to construct an MDP model as the abstract model (i.e., as the surrogate) of the AI-CPS to realize an efficient safety analysis.
The definition of MDP is given as follows.

\begin{mydefinition}[Markov Decision Process]
An MDP \mdp can be represented as a tuple (\mdpstate, \mdpinitalstate, \action, \mdptransition, \prob, \ap, \mdplabel) consisting of a finite set of states \mdpstate, an initial state \mdpinitalstate $\in \mdpstate$, a finite set of actions \action, a finite set of transitions \mdptransition~{$: \mdpstate \times \action \rightarrow \mdpstate$}, a transition probability function \prob~{$: \mdpstate \times \action \times \mdpstate \rightarrow [0,1]$}, a set of atomic propositions \ap, and a labeling function \mdplabel~{$:$ \mdpstate $\rightarrow 2^\ap$}.
\end{mydefinition}


The MDP model \mdp is derived based on a four-level abstraction that considers state, transition, action, and labeling.
We denote the state, transition, action, and labeling abstraction functions as \stateabst, \transabst, \actabst, \labelabst, respectively, the details of which are presented as follows.

\myparagraph{State Abstraction}
Given a Moore machine state $q$, the state abstraction function \stateabst maps it to an MDP concrete state $s$, i.e., \stateabst$(q) = s$.
The abstraction procedure contains two steps: \emph{automated dimension reduction} and \emph{equal interval partition}.
We first apply automated dimension reduction to the states $\moorestate$ of the Moore machine to resolve the problem of high dimensionality.
Specifically, this is achieved by employing principal component analysis (PCA) that transforms the Moore machine state $q \in \mathbb{R}^{l}$ to a low dimensional state $\hat{q} \in \mathbb{R}^{k}$ with $l > k$~\cite{du2019deepstellar}, which is used as the foundation of the subsequent abstractions. We denote the process of PCA as function $f$, and $\hat{q} = f(q)$.
Then, the $k$-dimensional reduced state space is partitioned into $\mathsf{c}^k$ regular grids~\cite{thompson1998handbook}, i.e., each dimension is equally partitioned into $\mathsf{c}$ intervals.
We denote the $i$-th interval on $j$-th dimension as $d^{j}_i$.
An MDP state $s$ thus contains the Moore machine states $\{q_1, \dots, q_n\}$ that fits in the same grid, i.e., $s = \{q_i | \hat{q}^1_i \in d^1_{\_} \bigwedge \dots \bigwedge \hat{q}^k_i \in d^k_{\_}, \hat{q}_i = f(q_i), i \in \{1, \dots,n\}\}$.



\myparagraph{Transition Abstraction}
The transitions between MDP states are obtained by the transition abstraction.
We use the transition abstraction function \transabst to map a transition of the Moore machine $\xi$ to an MDP transition $\theta$, i.e., \transabst$(\xi) = \theta$.
If there exists a Moore machine transition $\xi \in \mooretransition$ between $q \in s$ and $q' \in s'$, then an MDP transition is set up accordingly between the MDP states $s$ and $s'$.  
Namely, an MDP transition includes all the Moore machine transitions that share the same starting and destination MDP states. 

Moreover, to empower the probabilistic safety analysis for AI-CPS, we facilitate with transition probability $\rho(s, \mathsf{act}, s') \in \prob $ for each transition. 
It is calculated based on the number of Moore machine transitions from $q \in s$ to $q' \in s'$ with input $\sigma \in \mathsf{act}$ and the total number of outgoing transitions from $s$, i.e., we have
\begin{align*}
    \rho(s, \mathsf{act}, s') = \frac{|\{(q, \sigma, q') | q \in s, \sigma \in \mathsf{act}, q' \in s'\}|}{| \{(q, \sigma, \_) | q \in s, \sigma \in \mathsf{act}\}|}.
\end{align*}






\myparagraph{Action Abstraction}
The action abstract function \actabst is designed to transform the input of the Moore machine $\sigma$ to a corresponding low dimensional MDP action $\mathsf{act}$, i.e., \actabst$(\sigma) = \mathsf{act}$.
Considering computational efficiency, we use the round function, which takes the integer part of $\sigma$, as the abstraction of the input.
In other words, given an input of the Moore machine $\sigma \in$ \mooreinalphabet, the MDP action $\mathsf{act}$ is abstracted as $\mathsf{act} =$ \actabst $(\sigma) = round(\sigma)$.

\myparagraph{Labeling Abstraction}
Finally, we perform abstraction on the output of the Moore machine, which is mapped to the labeling of the MDP.
Recall that the output \mooreoutput is the robust semantics of the STL specification, which gives a continuous value in a way that a positive (negative) value represents the system is in the safe (unsafe) status.
Considering that an MDP state $s$ may contain multiple states $q_1 \dots q_n$, we define the labeling abstraction function as
\begin{align*}
    \zeta_L(\mooreoutput(q)) \;\Defeq\; 
    \begin{cases}
    -1 & \text{if } \min_{i = 1}^n \mooreoutput(q_n) < \varepsilon, q \in s=\{q_1,\dots,q_n\} \\
    +1 & \text{otherwise}
    \end{cases}
\end{align*}

Intuitively, if the minimum value of the output over $\{q_1, \dots q_n\}$ is smaller than a predefined threshold $\varepsilon$ or is negative, the system is close to or already in the dangerous status.
In such a case, we label the output of $s$ as $-1$, otherwise, we label it as $+1$.



\subsection{MDP Model Refinement}\label{subsec:MDPref}
For ensuring a reliable and accurate safety analysis based on the constructed MDP model, we further introduce a refinement procedure in this subsection.
The purpose of this refinement is to enhance the state abstraction function, such that the estimation error of the MDP model w.r.t. the state space, is reduced.

An overview of the refinement procedure is presented in Algorithm~\ref{alg:refinement}.
For each state $s$ of the input MDP \mdp, we first identify the output values $\mooreoutput(q)$ of all states $q$ that belong to it (Line 1-2).
Then, every state $q$ is classified into two sets according to this value, i.e., we add $q$ to $\mathcal{S}^{+}_{con}$ if $\mooreoutput(q) \geq 0$, and to $\mathcal{S}^{-}_{con}$ if $\mooreoutput(q) < 0$ (Line 3-4).
Thereafter, we compute a variance error $\frac{\sum_{q \in s} (\mooreoutput(q) - \overline{\mooreoutput(q)} )^2 }{|\mathcal{S}^{+}_{con} \cup \mathcal{S}^{-}_{con}|}$ over all states $q$ (Line 5).
If this variance error is larger than a predefined threshold $\varepsilon$, we train an SVM classifier to represent a new decision boundary for classifying states $q$ as safe and unsafe accordingly (Line 6).
Such an SVM classifier is adopted as the new state abstract function \stateabst (Line 7) and a refined MDP model is therefore obtained. 
The proposed refinement process is able to increase the preciseness of the abstract MDP model, which is used as the basis for an effective safety analysis for AI-CPSs. 




\begin{algorithm}[!tb]
	\caption{Refinement of our safety-analysis oriented MDP}
	\label{alg:refinement}
	\begin{algorithmic}[1] 
	\Require{an MDP \mdp = (\mdpstate, \mdpinitalstate, \action, \prob, \ap, \mdplabel), a robustness variance error threshold $\varepsilon$, an SA-Moore Machine \moore = (\moorestate, \mooreinitialstate, \mooreinalphabet, \mooreoutalphabet, \mooretransition, \mooreoutput), and the action/labeling/state/transition abstraction functions \actabst, \labelabst, \stateabst, \transabst.
	}
	\Statex
		\For{$s \in \mdpstate$}
		    \For{$q \in s$}
		        \State $\mathcal{S}^{+}_{con} \gets \mathcal{S}^{+}_{con} \cup q$ if $G(q) >= 0$ \Comment{G(q) is the robustness of the concrete states}
		        \State $\mathcal{S}^{-}_{con} \gets \mathcal{S}^{-}_{con} \cup q$ if $G(q) < 0$ 
		    \EndFor
            \If{$\frac{\sum_{q \in s} (G(q) - \overline{G(q)} )^2 }{|\mathcal{S}^{+}_{con} \cup \mathcal{S}^{-}_{con}|} > \varepsilon$ }
		            \State $\mathcal{B} \gets trainSVMClassfier( \mathcal{S}^{-}_{con}\cup \mathcal{S}^{+}_{con}, \mathcal{S}^{+}_{con}.labels, \mathcal{S}^{-}_{con}.labels)$
		            \State $\zeta_{\mathsf{S}} \gets ReconstructAbst$ ($\zeta_{\mathsf{S}}, \mathcal{B}$)
           \EndIf
		\EndFor
	    \State \mdp = \textit{reconstruct\_MDP}(\moore, \actabst, \labelabst, \stateabst, \transabst) \Comment{Reconstruct MDP with new abstraction functions}
	\end{algorithmic}
\end{algorithm}

\section{Online and Offline Safety Analysis}
\label{sec:application}

By leveraging the constructed MDP model as the abstract model of AI-CPS, it provides the possibility to further perform efficient safety analysis.
In this section, we introduce novel safety analysis techniques in two directions: online safety monitoring (Section~\ref{subsec:monitoring}) and offline model-guided falsification (Section~\ref{subsec:falsification}).
Our analysis methods initialize an early step to provide safety issue detection of AI-CPS and potentially increase the reliability of a deployed AI-CPS.


\subsection{Online Safety Monitoring}
\label{subsec:monitoring}
The goal of online safety monitoring is to provide safety suggestions (and potential countermeasures to safety hazards) during the real-time control process of the AI-CPS. 
To this end, we propose an efficient model-based safety monitoring technique that utilizes the derived MDP model to compute safety predictions. 


We leverage PMC as the basis for performing the online safety analysis.
As an automated verification technique, PMC focuses on providing not only Boolean results of the model checking but also quantitative guarantees for systems that exhibit probabilistic and randomized behaviors.
It adopts probabilistic computation tree logic (PCTL)~\cite{ciesinski2004probabilistic} as the foundation for the verification, which is defined as follows.

\begin{mydefinition}[Probabilistic Computation Tree Logic]
PCTL is a variant of temporal logic and is composed of state formula $\phi$ and path formula $\psi$, which are defined as 
\begin{align*}
    \phi \,::\equiv\,true \mid \alpha \mid \phi_1 \wedge \phi_2 \mid \neg \phi \mid \pctlProbOp{\sim p}[\psi] 
\end{align*}
\begin{align*}
    \psi \,::\equiv\, \NextOp \phi \mid \Box \phi \mid \Diamond \phi \mid \phi_1 \pctlUntilOp{\leq k} \phi_2 \mid \phi_1 \UntilOp{} \phi_2 \mid \mathcal{F}_{\leq k} \phi
\end{align*}
where $\pctlProbOp{}$ is the probabilistic operator, $a$ is an atomic proposition, $p \in [0,1]$ is the probability bound. 
We have $\sim \in \{<,\leq,>,\geq\}$ and $k \in \mathbb{N}$.
$s \models \pctlProbOp{\sim p}[\psi]$ means that the probability from state $s$, that $\psi$ is true for an outgoing path satisfies $\sim\xspace p$.
\end{mydefinition}

PCTL is designed to describe the behavior of a Markov model.
It supports not only the classic Boolean semantics, which is used for non-probabilistic operators but also the quantitative semantics designed for probabilistic operators.
By taking an MDP model \mdp, a state s $\in \mdpstate$, and a PCTL formula $\phi$ as the inputs, PMC outputs "verified" if $s \models \phi$, or otherwise, a counterexample (error path).

The online safety monitoring is then performed by employing PMC to compute the output of a given PCTL formula, which describes the safety status of the system and is referred to as the \emph{safety query} in this work, based on the constructed MDP model.
An example of the safety query could be $s \models \pctlProbOp{> 0.5}[\NextOp (rob=-1)]$, which means that "is the probability that for state $s$, the robustness is -1 in the next timestamp greater than 50\%?".
If the PMC returns "safe", we consider the system to be in a safe status.
Otherwise, the result indicates that the system is in an unsafe condition.





Based on the results of PMC, we further introduce a corresponding switching control strategy that switches between the AI controller and a predefined traditional safety controller for keeping the AI-CPS safe, in the commonly adopted safety redundancy contexts (see Fig.~\ref{fig:safetyMonitoring}). 
If the PMC outputs "unsafe", the traditional safety controller will be activated to ensure the safety of the system. 
Otherwise, the AI controller continues to determine the actually applied actions for increasing the overall efficiency of the AI-CPS~\cite{Song2021WhenCS,zhang2022falsifai}.
By using the proposed online safety monitoring technique, we can potentially design a hybrid control system that takes advantage of both the efficient AI controller and the traditional controller.


    



\begin{figure}[!t]
\centering
\includegraphics[width=0.4\textwidth]{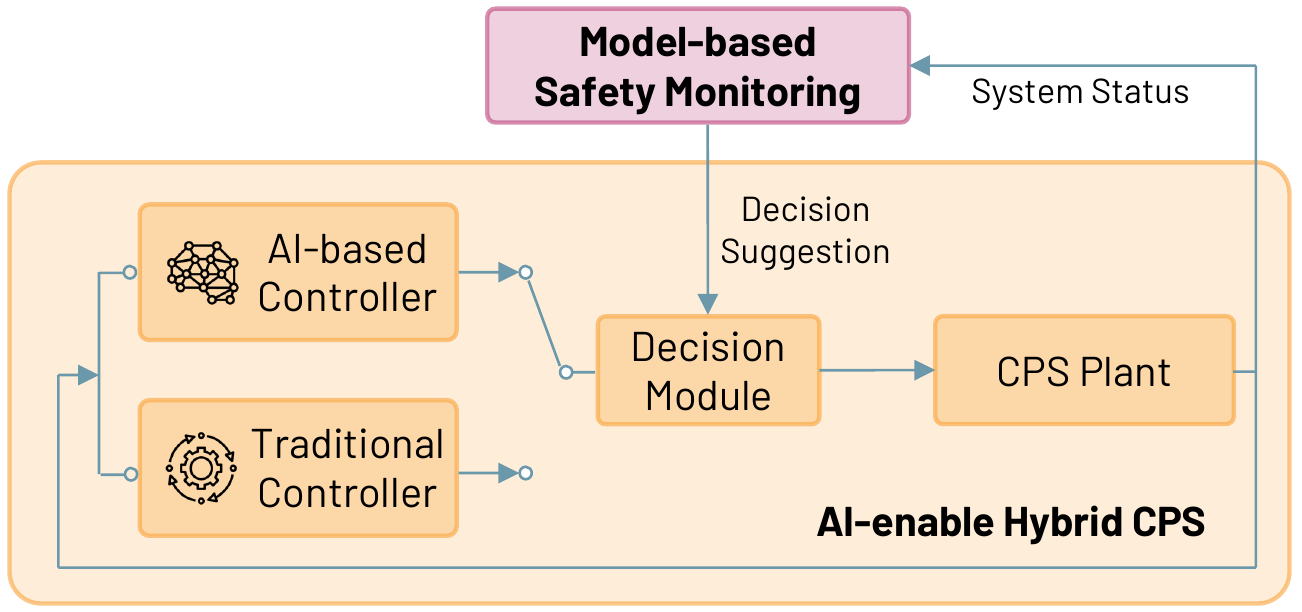}
\caption{Overview of the model-based safety monitoring}
\label{fig:safetyMonitoring}
\end{figure}
\vspace{-10pt}

\subsection{Offline Model-guided Falsification}
\label{subsec:falsification}

Falsification has been widely adopted for the safety assurance of CPS in detecting input cases that could trigger system behaviors that violate safety requirements. 
However, the existing optimization-based falsification techniques for traditional CPS are recently found to be ineffective for AI-CPS~\cite{Song2021WhenCS}, when the AI components are included as the key parts of the system.
Thus, as another promising direction of performing the safety analysis, in this subsection, we present a novel offline model-guided falsification method designed for AI-CPS. 

An overview of the proposed offline model-guided falsification is summarized in Fig.~\ref{fig:falsification}, which consists of two stages: a \emph{global} and a \emph{local} search stages.


A high-level description of our technique is as follows.
Specifically, we first adopt randomly generated input signals as candidate signals.
In the global search stage, for the candidate signal, which is generated randomly or based on the local search, we perform PMC on the abstract model with the signal, w.r.t. the safety query.
If the PMC returns "unsafe", which means the candidate could lead to an unsafe region, we put the signal into a queue for local search.
The local search process is realized by using stochastic optimization approaches, e.g., hill-climbing optimization~\cite{selman2006hill}, which are often known as effective in local exploitation.
The signal, returned by the local search, is then sent back to the global search procedure as a promising candidate signal.


Algorithm~\ref{alg:pmcFal} summarizes the detailed offline model-guided falsification process.
In particular, the inputs to the algorithm are the AI-CPS $\plant^\NN$ that is controlled by the AI controller $\NN$, the abstract MDP model $\mdp$, the desired STL specification $\varphi$ for the system, a PCTL specification $\phi$ for safety query, the initial size $k$ of the seed queue $\queue$, a global budget $t_g$, and a local budget $t_l$. 
The details of the algorithm are as follows:
\begin{compactitem}[$\bullet$]
\item First, \queue is initialized by sampling $k$ input signals randomly in the system input space (Line~\ref{line:initAlg2}), and the algorithm enters the outer loop (Line~\ref{line:outerLoop});
\item Then, the algorithm enters an inner loop (Line~\ref{line:innerLoop}), in which input signals are iteratively selected and evaluated. Specifically, the inner loop performs the following steps:
\begin{compactitem}[-]
\item at iteration $i$, an input signal $s_i$ is selected by: 
\begin{inparaenum}[i)]
\item popping out the head  of \queue if $i = 1$, or
\item running hill-climbing optimization \localOpt based on the sampling history of input signals, which plays the role of local search for falsifying inputs (Line~\ref{line:dequeueAndOpt});
\end{inparaenum}
\item the robustness $\rb{i}$ of the output signal $\plant^\NN(s_i)$ w.r.t. $\varphi$ is computed (Line~\ref{line:robustCalc});
\item the PMC of $s_i$ is conducted on $\mdp$ w.r.t. $\phi$ to decide whether it can lead to a potentially unsafe region (global search)(Line~\ref{line:pmc}); 
\item if $\rb{i}$ is negative, $s_i$ is returned as a falsifying input; otherwise, $s_i$ will be inserted to $\queue$ if it is possible to guide to a possible danger region, serving as guidance to future system behavior exploration (Line~\ref{line:robNeg}-\ref{line:enqueue});
\end{compactitem}
\item The algorithm can also terminate if no falsifying input is found within the global budget $t_g$ (Line~\ref{line:outerLoop}).
\end{compactitem}

Note that, the usage of the queue \queue as an auxiliary is due to its \emph{first-in-first-out} property; by that, a balance between \emph{exploration} and \emph{exploitation} is achieved, in the sense that a most-recently visited signal $s$ will be placed at the rear of \queue, and so other signals ahead of $s$ in \queue can be prioritized as the initial points for the local exploitation by the optimization. 
We select STL as the specification language for falsification since it is widely adopted in the falsification community~\cite{donze2010robust,Song2021WhenCS,zhang2022falsifai,zhang2020hybrid,zhang2021effective}.



\begin{figure}[!t]
\centering
\includegraphics[width=0.45\textwidth]{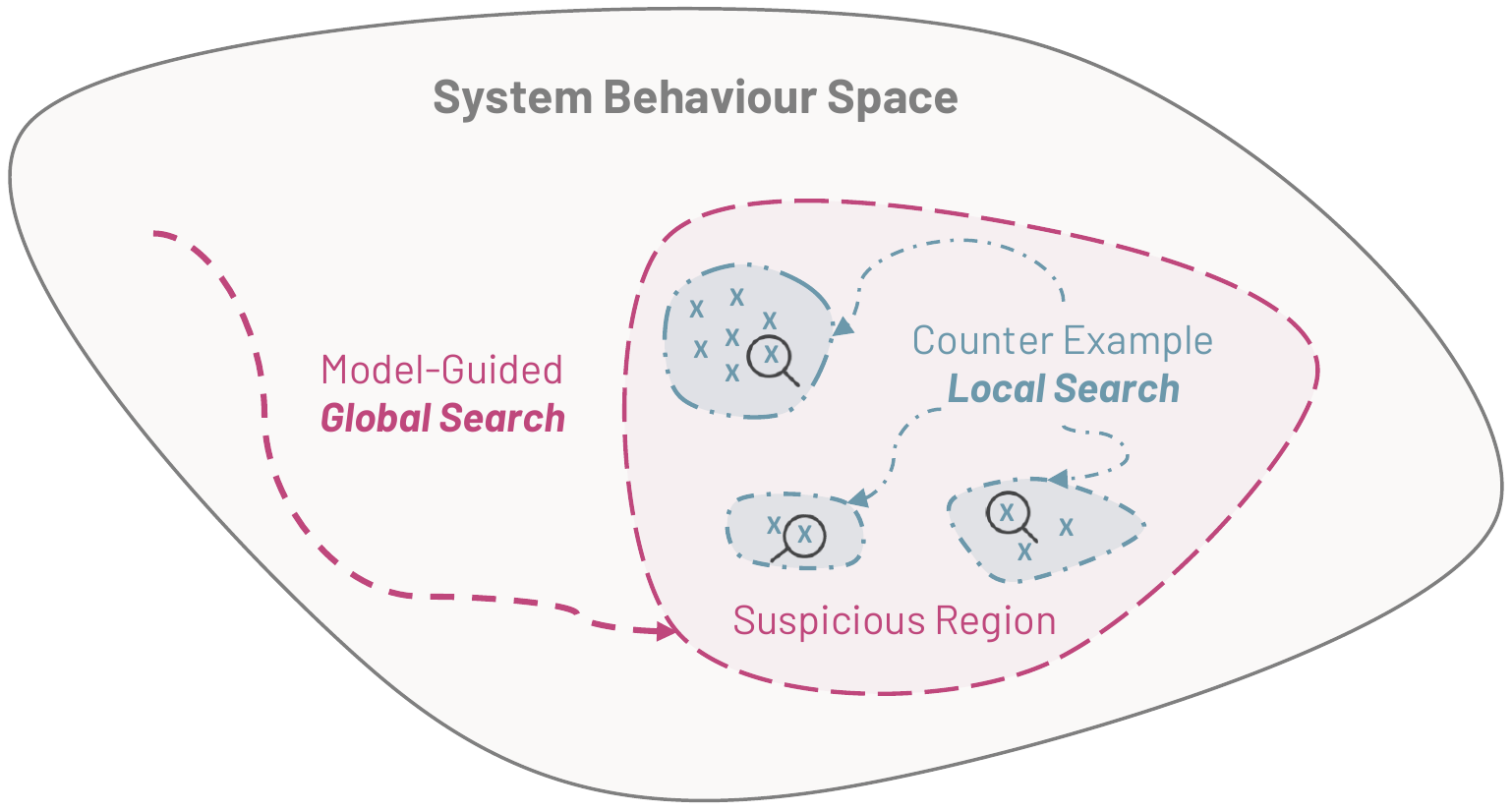}
\caption{Overview of model-guided falsification for AI-CPS}
\label{fig:falsification}
\end{figure}
\vspace{-10pt}

\begin{algorithm}[!tb]
	\caption{The offline model-guided falsification algorithm}
	\label{alg:pmcFal}
	\begin{algorithmic}[1] 
	\Require{
	an AI-CPS $\plant^\NN$ controlled by $\NN$, 
	the MDP $\mdp$, 
	an STL specification $\varphi$,
	a PCTL specification $\phi$,
	the initial size $k$ of the seed queue $\queue$, 
	a global budget $t_g$ and a local budget $t_l$. 
	}
	\Statex
		\State Initialize $t$ with 0, and $\queue$ with $k$ randomly sampled input signals \label{line:initAlg2}
		\While{$t < t_g$} \label{line:outerLoop}
		\State $t \gets t + 1$
		\For{$i \in \{1, ..., t_l\}$}\label{line:innerLoop}
		\State $x_i \gets \begin{cases}
	 \textsc{DeQueue}(\queue) & \text{if } i = 1 \\
		 \localOpt(\{x_l, \rb{l}\}_{l = 0, \ldots, i-1}) & \text{otherwise}
		\end{cases}$\label{line:dequeueAndOpt} \Comment{take an input from \queue or optimization for falsifying inputs (local search)}
		\State $\rb{i} \gets \Robust{\plant^\NN(x_i)}{\varphi}$\label{line:robustCalc} \Comment{check the robustness of $x_i$}
		\State $\flag{i} \gets \textsc{ProbabilisticMC}(
		x_i, \phi, \mdp
		)$ \label{line:pmc} \Comment{check if $x_i$ could lead to unsafe region (global search)}
		\If{$\rb{i} < 0$}\label{line:robNeg}
		\State \textbf{return} $x_i$ \Comment{$x_i$ is a falsifying input, return it}
		\ElsIf{\flag{i} is \textrm{true}} \label{line:novTrue}
		\State $\queue \gets \textsc{EnQueue}(\queue, x_i)$\label{line:enqueue} \Comment{$x_i$ may lead to falsification, push it to \queue}
		\EndIf
		\EndFor
		\EndWhile
	\end{algorithmic}
\end{algorithm}

\begin{table*}[!tb]
\centering
\caption{The specifications for falsification}
\label{tab:falspec}
\vspace{-10pt}
\resizebox{\textwidth}{!}{%
\begin{tabular}{lll}
\toprule
\textbf{Subject CPS} & \textbf{Specification} & \textbf{Description} \\\hline
{\acc} & \STLspec{\acc}{1} \;=\; 
\shortstack{$\Box_{[0,50]}(\mathtt{D_{rel}} \ge \mathtt{D_{safe}} + 1.4 * \mathtt{v_{ego}})$}
& \multirow{1}{*}{\shortstack{Retain a safe distance plus a braking distance of the ego car during the execution time}} \\

&\STLspec{\acc}{2} \;=\; 
\shortstack{$\Box_{[0,50]}
\begin{array}{l}
\left((\drel < \dsafe + 1.4*\vego) \to 
\Diamond_{[0,5]}(\drel > \dsafe + 1.4*\vego)\right)
\end{array}$} & Return to the safe status in 5 seconds if two cars are too closed\\
\hline

\afc & \STLspec{\afc}{1} \;=\; \shortstack{$\Box_{[0,30]}\left(\left\lvert\frac{\af - \afref}{\afref}\right\rvert < 0.1\right)$} & Maintain the air-to-fuel ratio in the reference value \\

 & \STLspec{\afc}{2}\;=\; \shortstack{$\Box_{[10,30]}\left(
(\left\lvert\frac{\af - \afref}{\afref}\right\rvert > 0.1)\to 
(\Diamond_{[0,1.5]}
\left\lvert\frac{\af - \afref}{\afref}\right\rvert < 0.1)
\right)$} & Recover to the reference value in 1.5 seconds if there is deviation of air-to-fuel ratio\\
\hline

\cstr & \STLspec{\cstr}{1} \;=\; \shortstack{$\Box_{[27,30]}\left(|\mathit{error}|  \le 0.35 \right)$} & Promote a reactor transition of conversion rate  \\

 \bottomrule
\end{tabular}
\vspace{-10pt}
}
\end{table*}

\section{Experimental Evaluation}
\label{sec:experiment}
To demonstrate the effectiveness, efficiency, and potential usefulness of \tool, we perform extensive evaluation and in-depth result analysis on representative CPSs.
In particular, we design experiments that aim to answer the following research questions (see also Section~\ref{subsec:overview}):
\begin{compactitem}[$\bullet$]
\item RQ1: How precise are the constructed abstract models?
\item RQ2: Can \tool provide effective safety monitoring?
\item RQ3: Is \tool effective in guiding the falsification procedure?
\item RQ4: How much overhead is introduced by the safety query in the analysis?
\end{compactitem}

Due to the page limit, in the rest of this section, we mainly discuss the summarized results on three representative CPSs (see CPS description in Section~\ref{subsec:benchmark}, the experimental setup in \ref{subsec:experimentsetup}, and results on each individual RQ in Section~\ref{subsec:rq1}-\ref{subsec:rq4}), while making the extensive and detailed evaluation results on more CPSs at our anonymous website.\footnote{\url{https://sites.google.com/view/ai-cps-mosaic}}







\subsection{Benchmark Systems}\label{subsec:benchmark}



\emph{Adaptive Cruise Control} (\acc), \emph{Abstract Fuel Control} (\afc), and \emph{Exothermic Chemical Reactor} (\cstr) are representative CPSs, which are widely used for CPS safety assurance research in previous work (e.g., \cite{jin2014powertrain, zhang2018two, ACCMathworks,Song2021WhenCS,zhang2021effective, zhang2022falsifai})


\myparagraph{Adaptive Cruise Control (\acc)}
The system is released by Mathworks~\cite{ACCMathworks}. 
This system targets to control the acceleration of the ego vehicle to keep the relative distance \drel to the lead car greater than a safety distance \dsafe. 
The safety distance is dynamically changed based on the relative velocity between two cars. 
In addition, when the safety distance is assured, the ego car should approach to use-set cruising velocity \vtarget.

\myparagraph{Abstract Fuel Control (\afc)}
\afc is released by Toyota~\cite{jin2014powertrain} which simulates a powertrain system of a vehicle. 
The external inputs are pedal angle and engine speed, and the fuel injection rate to a fuel cylinder is controlled by the system.
The air-fuel-ratio \af inside the cylinder is required to maintain a reference value \afref to achieve complete gasoline fuel combustion. 
The optimized stoichiometric air-fuel-ratio is about $14.7$~\cite{al2019analyzing} and the deviation should be limited within $0.1$.

\myparagraph{Exothermic Chemical Reactor (\cstr)}
\cstr is originally released by Mathworks~\cite{CSTRMathworks} which is a widely used system in various chemical industry domains. The concentration of the reagent in the exit stream of the reactor is expected to sustain the setpoint. When the reactor is triggered from a low transformation ratio to a high transformation ratio, the concentration setpoint will change accordingly. In addition, the error of the concentration should be less than $0.35$ during the entire transformation.




\begin{table}[!tb]
\centering
\caption{Abstract models details}
\label{tab:mdpstatetrans}
\vspace{-10pt}
\scriptsize
\setlength{\tabcolsep}{4pt}
{%
\begin{tabular}{cccccc}
\toprule
Benchmark & \acc-\ddpg & \acc-\sac & \acc-\tdthr & \acc-DNN$_1$ & \acc-DNN$_2$ \\\cline{2-6}
\#states & 376 & 1507 & 629 & 1091 & 337 \\
\#transitions & 1983 & 23396 & 12857 & 94574 & 2729 \\
\midrule
Benchmark & \afc-DNN$_1$ & \afc-DNN$_2$ & \cstr-\ddpg & \cstr-\tdthr &  \\\cline{2-6}
\#states & 783 & 367 & 724 & 323 &  \\
\#transitions & 2875 & 1208 & 9465 & 2087 & \\
\bottomrule
\vspace{-10pt}
\end{tabular}
}
\end{table}

\subsection{Experimental Setup}
\label{subsec:experimentsetup}
By considering the RQs that we plan to investigate, we design the evaluation experiments accordingly.
Details about the experimental setup are given as follows.

\myparagraph{RQ1}
First, we would like to investigate the preciseness of the abstract model in terms of labeling, which has potential influences on the subsequent safety analysis.
For this purpose, we randomly sample 1,000 traces for each derived abstract model. 
Then, for all concrete states $q$ contained in the trace, we check whether the output \mooreoutput($q$) (the robust semantics) matches the labeling of the corresponding abstract state $s$, i.e., we verify if \mdplabel$(s) =$ \labelabst(\mooreoutput($q$)), where $q \in s$.
The average percentage of the matched states is computed for evaluation.

\myparagraph{RQ2}
In this RQ, we examine whether \tool is able to improve the safety of AI-CPS while keeping a similar functional performance compared to the original system.
The corresponding safety and performance metrics used in the evaluation are given in Table~\ref{tab:RQ2eval}. 
We run the simulations of each AI-CPS with 100 
randomly generated signals and record the total number of time steps $t$ that satisfy the metrics in a simulation. 
The average percentage of safe and well-performed time steps over 100 simulations is computed for evaluation. 

\myparagraph{RQ3}
We also aim to investigate if \tool outperforms existing state-of-the-art falsification techniques for AI-CPSs. 
For this, we compare the performance of \tool with three falsification approaches, i.e., \texttt{Random}, \breach, and $\tool_{rand}$.
\texttt{Random} is a simple falsification method that randomly samples input signals from the input space, and feeds them to the system to run simulations.
\breach is a state-of-the-art falsification tool that employs the classic optimization-based falsification method~\cite{donze2010breach}.
It integrates multiple stochastic optimization solvers, e.g., CMA-ES and Simulated Annealing. 
We select CMA-ES as the solver for \breach in our experiments, as it has been reported as the most effective solver~\cite{zhang2018two}.
$\tool_{rand}$ is a variant of the proposed \tool, where the \textsc{enqueue} operation based on PMC (Line 11 in Algorithm~\ref{alg:pmcFal}) is replaced by random sampling in the input space.
We introduce $\tool_{rand}$ for better analyzing the effect of PMC in the falsification procedure. 

The effectiveness of different falsification techniques is evaluated by using the following metrics: 
\begin{inparaenum}[(i)]
\item falsification success rate ($\mathit{FSR}$): the number of runs, out of 30, that the falsification approach is successful, i.e., it finds a falsifying input for the given specification;
\item $\mathit{time}$: the average time of the successful falsification trials, i.e., a counterexample is detected;
\item $\mathit{\#sim}$: the average number of simulations of the successful falsification trials.
\end{inparaenum}

Based on these metrics, we perform comprehensive evaluations on the three selected benchmark systems.
In total, nine different AI controllers, which are explained in the following paragraphs, and five STL specifications, which are given in Table~\ref{tab:falspec}, are implemented for verifying the performance of \tool in depth.

\myparagraph{RQ4}
Lastly, we would like to evaluate the overhead from the safety monitoring process to the system and the impact of the time by the query. 
Therefore, for each AI-CPS, we run the simulation 100 times and examine the time spent on system simulation and the safety query respectively.

\myparagraph{Traditional \& AI Controllers}
We use model predictive control~\cite{ACCMathworks,CSTRMathworks} for \acc and \cstr, PI and feedforward controller~\cite{jin2014powertrain} for \afc.
As mentioned in Section~\ref{subsec:aicps}, we investigate both DRL and supervised learning-based AI controllers in this work. 
Therefore, for \acc, we perform experiments with five AI controllers:
\ddpg, \sac, \tdthr, and two DNN controllers with the following number of layers and hidden neurons: 3*50 (DNN$_1$) and 4*30 (DNN$_2$);
for \afc, we use two DNN controllers which have the following number of layers and hidden neurons: 3*15 (DNN$_1$) and 4*10 (DNN$_2$);
for \cstr, we mainly evaluate two DRL controllers, i.e., \ddpg and \tdthr.
In total, nine AI controllers are implemented in the experiments.


\myparagraph{Details of Abstract Models} 
The number of states and transitions of the abstract model is summarized in Table~\ref{tab:mdpstatetrans}.
For each abstract model, we collect 20,000 traces by using random sampling in the input space.
The state abstraction parameters $\mathsf{c}$ and $k$ are selected as 10 and 3, respectively. 
The refinement parameter $\varepsilon$ is $0$.
For the labeling of MDP, we choose \STLspec{\acc}{1}, \STLspec{\afc}{1}, and \STLspec{\cstr}{1}, such as shown in Table~\ref{tab:falspec}, for the three selected CPSs, which reflect the safety requirements of the system.

\myparagraph{Safety Query}
For the safety query, we use the following specification for the PMC on the constructed MDP.
\begin{align*}
s \models \pctlProbOp{> 0.8}[\mathcal{F}_{\leq 10} (rob=-1)]
\end{align*}
which means "\emph{Is the probability that for state $s$, the robustness is -1 in the future 10 steps, greater than 80\%?}".
Recall that we need to select a state for PMC.
For safety monitoring, at every five seconds of the simulation, we select the state $s$ of the current time step and then conduct PMC based on $s$.
For the offline falsification, we choose the state at the five second of the simulation, as the state for PMC.

\myparagraph{Software Dependencies \& Hardware Platforms}
We implement \tool in \matlab with the library of \simulink, and Python.
For PMC, we adopt \prism~\cite{kwiatkowska2011prism} as the verification engine.
The falsification algorithm of \tool is implemented on top of \breach~\cite{donze2010breach} \textsc{1.9.0}, which is also considered as one baseline approach for falsification in our comparisons.
The \acc model requires the \emph{Model Predictive Control} and \emph{Control System} \matlab toolboxes.

For the abstract MDP model construction and refinement, we run the approaches on two servers, each with Intel(R) Core(TM) i9-10940X CPU @ 3.30GHz Processor, 28 CPUs, 62G RAM with two NVIDIA RTX A6000. 
For collecting sampling traces, falsification trials,
the safety monitoring evaluation, and the overhead analysis, we run the proposed framework with a Lambda Tensorbook, which is equipped with an Intel(R) Core(TM) i7-11800H @ 2.30GHz Processor with 8 CPUs and 64G RAM, and an NVIDIA RTX 3080 Max-Q GPU with 16 GB VRAM.



%
\begin{table}[!tb]
\centering
\caption{The STL specifications for evaluating safety monitoring. 
Safety follows the pattern: $\Box_I(\varphi_1)$; Performance follows the pattern: $\Box_I(\varphi_2)$;
}
\label{tab:RQ2eval}
\vspace{-10pt}
\scriptsize
\setlength{\tabcolsep}{4pt}
\begin{tabular}{ccc}
\toprule
 Systems & Safety & Performance \\\hline
\multirow{3}{*}{\acc} 
& \multirow{3}{*}{\shortstack{$I = [0,50]$\\$\varphi_1 \equiv \drel \ge \dsafe$}} 
& \multirow{3}{*}{\shortstack{$I = [0,50]$ \\ 
$\varphi_2 \equiv | \vego - \vtarget | \le 0.2$
}}
\\ \\ 
 \\ \hline


\multirow{3}{*}{\afc} 
& \multirow{3}{*}{\shortstack{$I=[0,30]$\\$\varphi_1\equiv \mu\le 0.02$}}  
& \multirow{3}{*}{\shortstack{$I=[0,30]$\\$\varphi_2\equiv \mu \le 0.016$}}  
\\
 &  &    \\ \\
 \hline

 \multirow{3}{*}{\cstr} 
& \multirow{3}{*}{\shortstack{$I = [25,30]$\\$\varphi_1 \equiv |\mathit{error}| \le 0.3$}}
& \multirow{3}{*}{\shortstack{$I = [25,30]$\\$\varphi_2 \equiv |\mathit{error}| = 0.24$}} \\
 &  &    \\ \\

 \bottomrule
 \vspace{-10pt}
\end{tabular}
\end{table}



\subsection{RQ1. The preciseness of Abstract Model}
\label{subsec:rq1}


Table~\ref{tab:rq1} presents the experimental results of evaluating the preciseness of the abstract model.
The average preciseness over all three benchmark systems is $92.22\%$. 
And for \acc, \afc, and \cstr, the average preciseness is $86.04\%$, $99.99\%$, and $99.91\%$, respectively.
It can be observed that \acc has a lower preciseness compared to the other two systems, which both have a value above $99.9\%$.
The reason for this could be that, the system state space of \acc has a higher dimension than \afc and \cstr, which requires more tuning on the abstraction parameters, e.g. $\mathsf{c}$ and $k$.


\begin{table}[!tb]
\centering
\caption{RQ1 -- The preciseness of the abstract models}
\label{tab:rq1}
\vspace{-10pt}
\scriptsize
\setlength{\tabcolsep}{4pt}
{%
\begin{tabular}{cccccc}
\toprule
Benchmark & \acc-\ddpg & \acc-\sac & \acc-\tdthr & \acc-DNN$_1$ & \acc-DNN$_2$ \\\cline{2-6}
Preciseness & 88.51\% & 91.31\% & 83.27\% & 84.15\% & 82.96\% \\
\midrule
Benchmark & \afc-DNN$_1$ & \afc-DNN$_2$ & \cstr-\ddpg & \cstr-\tdthr &  \\\cline{2-6}
\#Preciseness & 99.99\% & 100\% & 99.85\% & 99.96\% &  \\
\bottomrule
\vspace{-10pt}
\end{tabular}
}
\end{table}

\begin{tcolorbox}[size=title]
	{\textbf{Answer to RQ1:}}
The constructed abstract model can precisely capture the labeling characteristics. 
\end{tcolorbox}
\vspace{-.1in}
\begin{table}[!tb]
\centering
\caption{RQ2 -- Experiment results for the online safety monitoring}
\label{tab:rq2_result}
\vspace{-10pt}
\scriptsize
\setlength{\tabcolsep}{4pt}
\resizebox{0.7\columnwidth}{!}{%
\begin{tabular}{cccc}
\toprule
Benchmark & Controller & Safety & Performance  \\
\midrule
\multirow{2}{*}{\acc-\ddpg} & \cellcolor{lightgray}{AI} & \cellcolor{lightgray}{0.9880} & \cellcolor{lightgray}{0.2934}  \\
& \cellcolor{lightgray}{\tool} & \cellcolor{lightgray}{0.9880} & \cellcolor{lightgray}{0.3134}  \\
\multirow{2}{*}{\acc-\sac} & AI & 0.9299 & 0.9390  \\
& \tool & 0.9559 & 0.9435 \\
\multirow{2}{*}{\acc-\tdthr} & \cellcolor{lightgray}{AI} & \cellcolor{lightgray}{0.8537} & \cellcolor{lightgray}{0.9418}  \\
& \cellcolor{lightgray}{\tool} & \cellcolor{lightgray}{0.9242} & \cellcolor{lightgray}{0.9606} \\
\multirow{2}{*}{\acc-DNN$_1$} & AI & 0.5109	& 0.9101 \\
& \tool & 0.9753 & 0.6363  \\
\multirow{2}{*}{\acc-DNN$_2$} & \cellcolor{lightgray}{AI} & \cellcolor{lightgray}{0.5311} & \cellcolor{lightgray}{0.8157} \\
& \cellcolor{lightgray}{\tool} & \cellcolor{lightgray}{0.9733} & \cellcolor{lightgray}{0.6924} \\
\multirow{2}{*}{\afc-DNN$_1$} & AI & 0.8449 & 0.7700   \\
& \tool & 0.7589 & 0.6658  \\
\multirow{2}{*}{\afc-DNN$_2$} & \cellcolor{lightgray}{AI} & \cellcolor{lightgray}{0.7863} & \cellcolor{lightgray}{0.6896} \\
& \cellcolor{lightgray}{\tool} & \cellcolor{lightgray}{0.7001} & \cellcolor{lightgray}{0.6038} \\
\multirow{2}{*}{\cstr-\ddpg} & AI & 0.8361 & 0.6988   \\
& \tool & 1 & 1   \\
\multirow{2}{*}{\cstr-\tdthr} & \cellcolor{lightgray}{AI} & \cellcolor{lightgray}{1} & \cellcolor{lightgray}{0.9643} \\
& \cellcolor{lightgray}{\tool} & \cellcolor{lightgray}{1} & \cellcolor{lightgray}{1}  \\
\bottomrule
\end{tabular}
}
\end{table}

\begin{table*}[!tb]
\caption{RQ3 -- Experiment result for falsification trials of \acc, \afc, and \cstr with their specifications. (\textit{FSR}: the number of successful falsification trials that found falsifying inputs (out of 30). \textit{time}: average time cost of successful trials. \textit{$\#$sim}: average number of simulations of successful trials) We highlight the best results in gray.}
\vspace{-10pt}
\label{tab:rq3falsification}
\scriptsize
\setlength{\tabcolsep}{5pt}
\begin{subtable}{\textwidth}
\resizebox{\textwidth}{!}{%
\begin{tabular}{crrr|rrr|rrr|rrr}
\hline
         & \multicolumn{3}{c}{\acc-\ddpg-\STLspec{\acc}{1}} & \multicolumn{3}{c}{\acc-\sac-\STLspec{\acc}{1}} & \multicolumn{3}{c}{\acc-\tdthr-\STLspec{\acc}{1}} & \multicolumn{3}{c}{\acc-DNN$_1$-\STLspec{\acc}{1}}  \\
         Alg. &    FSR     &    time    &    \#sim    &    FSR    &   time     &   \#sim     
         &   FSR  & time  &   \#sim    
         &   FSR   &  time     &   \#sim           \\\hline
\texttt{Random}
& 0	& - & - 
& 6 & 90.4 & 71.7     
& 10 & 95.9 & 47.0  
& 5 & 368.5 & 83   \\
\breach 
& 0 & - & -
& 9 & 65.6 & 51.7 
& 18 & 39.0 & 34.1
& 0 & - & -  \\
$\tool_{rand}$
& \cellcolor[gray]{0.8} 9 & \cellcolor[gray]{0.8} 560.4 & \cellcolor[gray]{0.8} 90.6
& 15 & 248.5 & 67.8    
& 11 & 307.6 & 87.2 
& 10 & 169.6 & 78.6   \\
\tool
& 4 & 714.2 & 108.5
& \cellcolor[gray]{0.8} 15 & \cellcolor[gray]{0.8} 244.5 & \cellcolor[gray]{0.8} 64.6
& \cellcolor[gray]{0.8} 22 & \cellcolor[gray]{0.8} 280.8 & \cellcolor[gray]{0.8} 49.8
& \cellcolor[gray]{0.8} 14 & \cellcolor[gray]{0.8} 268.3 & \cellcolor[gray]{0.8} 51.1 \\
\hline
\end{tabular}
}
\end{subtable}
\vfill
\begin{subtable}{\textwidth}
\resizebox{\textwidth}{!}{%
\begin{tabular}{crrr|rrr|rrr|rrr}
\hline
         & \multicolumn{3}{c}{\acc-DNN$_2$-\STLspec{\acc}{1}} & \multicolumn{3}{c}{\acc-\ddpg-\STLspec{\acc}{2}} & \multicolumn{3}{c}{\acc-\sac-\STLspec{\acc}{2}} & \multicolumn{3}{c}{\acc-\tdthr-\STLspec{\acc}{2}}  \\
         Alg. &    FSR     &    time    &    \#sim    &    FSR    &   time     &   \#sim     
         &   FSR  & time  &   \#sim    
         &   FSR   &  time     &   \#sim           \\\hline
\texttt{Random}
& 2 & 370.0 & 70 
& 0 & - & -   
& 0 & - & -
& 1 & 49.7 & 25.0 \\
\breach 
& 4 & 192.3 & 41.8
& 8 & 143.0 & 86.8 
& \cellcolor[gray]{0.8} 12 & \cellcolor[gray]{0.8} 95.7 & \cellcolor[gray]{0.8} 72.9
& 14 & 88.9 & 50.1 \\
$\tool_{rand}$
& 19 & 160.4 & 60.7 
& 4	& 203.2 & 132.5    
& 7	& 197.4 & 113.7  
& 10 & 105.9 & 70.6\\
\tool
& \cellcolor[gray]{0.8} 20 & \cellcolor[gray]{0.8} 251.3 &  \cellcolor[gray]{0.8} 47.0
& \cellcolor[gray]{0.8} 9 & \cellcolor[gray]{0.8} 560.4 & \cellcolor[gray]{0.8} 90.6
& 11 & 307.6 & 87.2 
& \cellcolor[gray]{0.8}16 & \cellcolor[gray]{0.8} 248.5 & \cellcolor[gray]{0.8} 67.8 \\
\hline
\end{tabular}
}
\end{subtable}
\vfill
\begin{subtable}{\textwidth}
\resizebox{\textwidth}{!}{%
\begin{tabular}{crrr|rrr|rrr|rrr}
\hline
         & \multicolumn{3}{c}{\acc-DNN$_1$-\STLspec{\acc}{2}} & \multicolumn{3}{c}{\acc-DNN$_2$-\STLspec{\acc}{2}} & \multicolumn{3}{c}{\afc-DNN$_1$-\STLspec{\afc}{1}} & \multicolumn{3}{c}{\afc-DNN$_2$-\STLspec{\afc}{1}}  \\
         Alg. &    FSR     &    time    &    \#sim    &    FSR    &   time     &   \#sim     
         &   FSR  & time  &   \#sim    
         &   FSR   &  time     &   \#sim           \\\hline
\texttt{Random}
& 0 & - & - 
& 0 & - & -   
& 6	& 136.6 & 50.3 
& 8 & 232.5 & 85.8     \\
\breach 
& 0 & - & -
& 21 & 245.7 & 56.6 
& 7 & 323.6 & 105.8 
& 6	& 553.5 & 212.5   \\
$\tool_{rand}$
& 9	& 195.4 & 71.5 
& 18 & 185.6 & 59.1    
& \cellcolor[gray]{0.8} 19 & \cellcolor[gray]{0.8} 190.7 & \cellcolor[gray]{0.8} 71.2 
& 10 & 199.3 & 76.7 \\
\tool
& \cellcolor[gray]{0.8} 11 & \cellcolor[gray]{0.8} 249.7  & \cellcolor[gray]{0.8} 70
& \cellcolor[gray]{0.8} 22 & \cellcolor[gray]{0.8} 206.5 & \cellcolor[gray]{0.8} 59.5 
& 12 & 326.6 & 54.5
& \cellcolor[gray]{0.8} 15 & \cellcolor[gray]{0.8} 422.3 & \cellcolor[gray]{0.8} 62.8  \\
\hline
\end{tabular}
}
\end{subtable}
\vfill
\begin{subtable}{\textwidth}
\resizebox{\textwidth}{!}{%
\begin{tabular}{crrr|rrr|rrr|rrr}
\hline
         & \multicolumn{3}{c}{\afc-DNN$_1$-\STLspec{\afc}{2}} & \multicolumn{3}{c}{\afc-DNN$_2$-\STLspec{\afc}{2}} & \multicolumn{3}{c}{\cstr-\ddpg} & \multicolumn{3}{c}{\cstr-\tdthr}  \\
         Alg. &    FSR     &    time    &    \#sim    &    FSR    &   time     &   \#sim     
         &   FSR  & time  &   \#sim    
         &   FSR   &  time     &   \#sim           \\\hline
\texttt{Random}
& 0 & - & - 
& 0 & - & -   
& 30 & 34.2 & 23.8  
& 1 & 141.7 & 71.0   \\
\breach 
& 4 & 673.2 & 186.0
& 0 & - & - 
& 18 & 64.8 & 46.5
& 1 & 35.6 & 31.0  \\
$\tool_{rand}$
& 3	& 98.8 & 45.0 
& 0	& - & -    
& 30 & 38.6 & 13.3 
& 12 & 110.9 & 60.1 \\
\tool
& \cellcolor[gray]{0.8} 4 & \cellcolor[gray]{0.8} 186.4 & \cellcolor[gray]{0.8} 26.2
& \cellcolor[gray]{0.8} 1 & \cellcolor[gray]{0.8} 277.7 & \cellcolor[gray]{0.8} 53.0 
& \cellcolor[gray]{0.8} 30 & \cellcolor[gray]{0.8} 44.2 & \cellcolor[gray]{0.8} 9.1
& \cellcolor[gray]{0.8} 16 & \cellcolor[gray]{0.8} 236.7 & \cellcolor[gray]{0.8} 57.4 \\
\hline
\end{tabular}
}
\end{subtable}
\end{table*}

\subsection{RQ2. Online Safety Monitoring}
\label{subsec:rq2}

Table~\ref{tab:rq2_result} shows the results of the analysis of online safety monitoring, where three key observations are identified.
First, \tool is able to improve the safety of the system while keeping a similar functional performance.
It can be observed that, in 5 experiments, both the safety and performance are increased.
Second, for \acc with DNN$_1$ and DNN$_2$, we notice that although the safety is indeed increased with the proposed online safety monitoring, the performance decreases.
The safety and performance metrics of the traditional controller, DNN$_1$, and DNN$_2$ are 0.9744, 0.5109, 0.5311 and 0.6475, 0.9101, 0.8157, respectively.
In other words, the AI-based controllers lean towards the performance metrics while failing at the safety of the system, while the traditional controller possesses an opposite control logic.
One reason for this could be that the used safety query is more prone to safety metrics.
Thus when switching to the traditional controller for safety, the performance of the system inevitably decreases.
Third, for \afc with DNN$_1$ and DNN$_2$, \tool performs worse than AI controllers.
The main reason could be that traditional controllers have worse safety and performance compared to AI controllers.
The safety and performance with traditional PI and feedforward controllers are 0.7177 and 0.6143, which are lower than \afc with DNN$_1$ and DNN$_2$.
In the case of a worse traditional controller, the switching control strategy could not help to improve the safety of the system.

\begin{tcolorbox}[size=title]
	{\textbf{Answer to RQ2:}}
\tool is effective in providing online safety monitoring while keeping a similar performance as the original AI-controlled system. 
In certain cases, the safety or performance decreases, which is mainly due to the limitation of traditional controllers or the safety query.
\end{tcolorbox}
\vspace{-.1in}

\subsection{RQ3. Falsification}
\label{subsec:rq3}

The experimental results of falsification are presented in Table~\ref{tab:rq3falsification}.
We run 30 falsification trials for each falsification approach and report the number of successful trials, i.e., $\mathit{FSR}$, as the indicator for the effectiveness of the approach. 
We highlight in the table the best approach in each benchmark system, where the first and second priorities are given to FSR and \#sim, respectively.
Based on the results, we can observe that: 
\begin{compactitem}[$\bullet$]
\item First, \tool obviously outperforms the other three falsification techniques.
In 13 out of 16 falsification problems, \tool has the best performance.
There are 9 trials where \tool is strictly better than all other approaches, i.e., the FSR is strictly great than other methods.
\item There are some experiments, e.g., \acc-\ddpg with \STLspec{1}{\acc} and \afc-DNN$_1$ with \STLspec{2}{\afc}, where \tool does not perform as well as $\toolRand$.
One possible reason for this is that, compared to $\toolRand$, \tool has a better balance between exploration and exploitation. 
Therefore, it spends more time in searching for suspicious regions, leading to a worse performance than the random exploration with optimization-based exploitation.
There is also one case where \breach performs the best (\acc-\sac with \STLspec{2}{\acc}), but only with one more found falsifying input compared to \tool.
\item \falRand can not outperform any other methods in the falsification trials. 
This means \falRand is the worst falsification approach, as it only conducts random exploration in the input space.
\end{compactitem}

\begin{tcolorbox}[size=title]
    {\textbf{Answer to RQ3:}}
\tool obviously outperforms the other three falsification methods, i.e., \breach, $\toolRand$, and \falRand.
There are some instances in which \breach or $\toolRand$ perform better than \tool.
\end{tcolorbox}
\vspace{-.1in}




\subsection{RQ4. Overhead Analysis}
\label{subsec:rq4}

\begin{table}[!tb]
\centering
\caption{RQ4 -- Experiment results for the overhead analysis. The time of the whole simulation and the safety query, and the ratio. (Time in seconds.)}
\label{tab:rq4_result}
\vspace{-10pt}
\scriptsize
\setlength{\tabcolsep}{4pt}
\resizebox{0.8\columnwidth}{!}{%
\begin{tabular}{cccc}
\toprule
Benchmark & Simulation & Safety Query & Ratio \\
\midrule
\acc-\ddpg & 7.7120 & 0.0990 & \cellcolor[gray]{0.8} 1.28\%   \\
\acc-\sac & 12.0166 & 0.0971 & \cellcolor[gray]{0.8} 0.81\% \\
\acc-\tdthr & 11.3914 & 0.0973 & \cellcolor[gray]{0.8} 0.85\% \\
\acc-DNN$_1$ & 27.9515 & 0.1156 & \cellcolor[gray]{0.8} 0.41\% \\
\acc-DNN$_2$ & 27.8731 & 0.1157 & \cellcolor[gray]{0.8} 0.42\% \\
\afc-DNN$_1$ & 155.5552 & 0.1017 & \cellcolor[gray]{0.8} 0.06\% \\
\afc-DNN$_2$ & 71.6134 & 0.1006 & \cellcolor[gray]{0.8} 0.14\%  \\
\cstr-\ddpg & 4.2881 & 0.1013 & \cellcolor[gray]{0.8} 2.36\%  \\
\cstr-\tdthr & 3.3493 & 0.1013 & \cellcolor[gray]{0.8} 3.02\% \\
\bottomrule
\end{tabular}
}
\end{table}

The results of the overhead analysis are shown in Table~\ref{tab:rq4_result}.
According to the results, it can be concluded that the overhead of the safety query for the online safety monitoring is almost negligible since the time spent on PMC is much less than in simulation.
In particular, for \afc-DNN$_1$ the differences between the average safety query time and the average simulation time are four orders of magnitude, and it is three orders of magnitude for \afc-DNN$_2$, \acc-\sac, and \acc-\tdthr.
The average simulation time over all benchmark systems is 35.7501 seconds, while the average query time is 0.1032 seconds, which is of two orders of magnitude difference.




\begin{tcolorbox}[size=title]
	{\textbf{Answer to RQ4:}}
Compared to the time of system simulation, the time spent on the safety query is negligible, which means \tool has little overhead.
\end{tcolorbox}
\vspace{-.1in}

\section{Threats to Validity}
\label{sec:threat2validity}

\myparagraph{External Validity}
Diverse system behaviors and various operation requirements among different AI-CPSs can be an external factor that impacts the validity of our results, since the proposed safety analysis method may not be effective on other CPSs. 
To mitigate this threat, we select CPSs from diverse and representative industry domains with different control requirements and operation specifications.

\myparagraph{Internal Validity} One potential threat is that CPSs can have varying performance w.r.t. different environmental parameters, especially when controlled by AI controllers. 
To mitigate such threats and keep consistency throughout the entire experiment, we use the same parameters for sample collection, model checking, falsification and evaluation. 
Moreover, we configure the environment setting and confirm the system behaviors with reference to the source documentation and related research works.

\myparagraph{Construct Validity} The evaluation metrics we used may not fully reflect the effectiveness of our approach. 
To mitigate this threat and more comprehensively assess our safety analysis methods, for each system, we leverage multiple specifications for falsification and safety monitoring from different aspects to deliver a comprehensive understating of the performance of our method. 

\section{Related Work}
\label{sec:relatedwork}

\myparagraph{Safety Analysis for AI-enabled Cyber-Physical System}
Quality assurance of AI-CPSs is of urgent need since many of its applications are in safety-critical domains. Therefore, there comes an increasing trend of relevant research activities in this direction recently~\cite{zolfagharian2022search, zhang2022falsifai,Song2021WhenCS,al2021robustness}.
For CPS with AI components, some recent works have also been proposed to assure its safety and reliability, e.g., Al-Nima et al.~\cite{al2021robustness} propose a Genetic Algorithm of Neuron Coverage (GANC) to improve the performance and robustness of DRL controllers in self-driving applications. 
Zolfagharian et al.~\cite{zolfagharian2022search} leverage ML models and genetic algorithms to test the reliability of DRL agents towards faulty episodes. 
In addition to offline testing for AI systems, our work provides a general safety analysis framework that also extensively provides online safety operation advice by safety monitoring. 
Our framework not only provides AI-aware guidance to a two-stage falsification but also empowers the performance of AI controllers during the execution phase. 

\myparagraph{Abstraction-based Analysis of AI systems}
Building an abstract model for DNNs recently attracted researchers' interest to facilitate the explainability of the AI model.
For example, Wang et al.~\cite{wang2018automatically} propose a method that applies learning and abstraction on system log traces to automatically enable formal verification of discrete-time systems.
Later, Du et al.~\cite{du2020marble} proposed MARBLE, a quantitative adversarial robustness analysis technique for recurrent neural networks (RNNs).
It extracts an abstract model from RNN to quantitatively measure the robustness information of RNN, which shows its effectiveness in the detection of adversarial examples.
In contrast, our work focuses on constructing a general-purpose safety analysis framework, and performing the safety guidance and detection for AI-CPS, which is more on the system level, where DNNs behave as the key components in the system.




\section{Conclusion}
\label{sec:conclusion} 

In this work, we present \tool, a model-based safety analysis framework for AI-CPS.
\tool first abstracts the system as an MDP, which is a representative to empower effective safety analysis.
With the abstract model, we further design two lines of safety analysis: online safety monitoring and offline model-guided falsification.
The evaluation demonstrates that \tool is effective in performing safety monitoring and finding falsifying inputs with neglectable overhead.
Based on \tool, in the future, we plan to perform more extensive research on safety assurance of AI-CPSs (e.g., fault localization, debugging and repairing), towards providing a general safety assurance framework to construct trustworthy AI-CPSs.




\balance

\bibliographystyle{ACM-Reference-Format}
\bibliography{main}










\end{document}